\documentclass[amssymb,prc,twocolumn,superscriptaddress,showpacs,aps]{revtex4}
\usepackage{graphicx}  
\usepackage{dcolumn}   
\usepackage{amsmath,bm}

\newcommand{\VAP}{VAPNP}

\begin{document}
\title{Variation after Particle-Number Projection for the HFB Method
with the Skyrme Energy Density Functional}

\author{M.V.~Stoitsov}
\affiliation{Department of Physics \& Astronomy, University of
Tennessee, Knoxville, Tennessee 37996, USA} \affiliation{Physics
Division,  Oak Ridge National Laboratory, P.O. Box 2008, Oak
Ridge, Tennessee 37831, USA} \affiliation{Joint Institute for
Heavy-Ion Research, Oak Ridge, Tennessee 37831, USA}
\affiliation{Institute of Nuclear Research  and Nuclear Energy,
Bulgarian Academy of Sciences, Sofia-1784, Bulgaria}

\author{J.~Dobaczewski}
\affiliation{Institute of Theoretical Physics, Warsaw University,
ul. Ho\.za 69, 00-681 Warsaw, Poland}

\author{R.~Kirchner}
\affiliation{Technische Universit\"at Wien, Karlsplatz 13,
A-1040 Wien, Austria}

\author{W.~Nazarewicz}
\affiliation{Department of Physics \& Astronomy, University of
Tennessee, Knoxville, Tennessee 37996, USA} \affiliation{Physics
Division,  Oak Ridge National Laboratory, P.O. Box 2008, Oak
Ridge, Tennessee 37831, USA} \affiliation{Joint Institute for
Heavy-Ion Research, Oak Ridge, Tennessee 37831, USA}
\affiliation{Institute of Theoretical Physics, Warsaw University,
ul. Ho\.za 69, 00-681 Warsaw, Poland}

\author{J.~Terasaki}
\affiliation{Department of Physics \& Astronomy, University of
Tennessee, Knoxville, Tennessee 37996, USA} \affiliation{Physics
Division,  Oak Ridge National Laboratory, P.O. Box 2008, Oak
Ridge, Tennessee 37831, USA} \affiliation{Joint Institute for
Heavy-Ion Research, Oak Ridge, Tennessee 37831, USA}

\date{\today}

\begin{abstract}
Variation after particle-number restoration is
incorporated  for the first time into the Hartree-Fock-Bogoliubov
framework employing  the Skyrme energy density
functional with
zero-range  pairing. The resulting projected HFB equations can be expressed
in terms of the local gauge-angle-dependent densities.
Results of projected calculations are compared with those obtained
within the Lipkin-Nogami method in the standard version and with
the Lipkin-Nogami method
followed by exact
particle-number projection.
\end{abstract}


\maketitle

\section{Introduction}
\label{sec0}

Pairing correlations affect
properties of atomic nuclei in a profound way
\cite{[Boh75],[RS80],[Bri05],[Dea03]}. They impact nuclear binding,
properties of nuclear excitations and decays, and dramatically
influence the nuclear collective motion.
In particular, pairing plays a crucial role in exotic, weakly bound nuclei
in which the magnitude  of the chemical potential is  close to that of
the pairing gap. In such systems, a naive independent single-particle
picture breaks down and the pair scattering, also involving the
continuum part of the phase space, can determine
the very nuclear existence \cite{[Dob98c]}.

Many aspects of
nuclear superfluidity can be successfully treated within the
independent  quasiparticle
framework by applying the
Bardeen-Cooper-Schrieffer (BCS) \cite{[Bar57]} or Hartree-Fock-Bogoliubov (HFB)
approximations \cite{[RS80]}.
The advantage of the mean-field approach to the pairing
problem lies in  its simplicity that allows for a straightforward
interpretation
in terms of pairing fields and deformations (pairing gaps)
associated with the spontaneous breaking of the gauge symmetry.
However, this simplicity comes at a cost. In
the intrinsic-system description,  the gauge angle associated with the
particle-number operator is fixed; hence,
the particle-number invariance is internally broken
\cite{[Boh75],[RS80],[Bri05]}. Therefore, to relate to
experiment, the  particle-number symmetry 
needs to be, in principle, restored.

Some observables, like masses, radii, or deformations are not very
strongly affected by the particle-number-symmetry restoration, while
some other ones, like even-odd mass staggering or pair-transfer amplitudes 
are influenced significantly. Moreover, quantitative impact of the
particle-number projection (PNP) depends on whether the pairing
correlations are strong (open-shell systems) or weak (near closed
shells). Therefore, methods of restoring the particle-number
symmetry must be implemented in studies of pairing correlations.
This can
be done on various levels \cite{[RS80],[Flo97]},
including the quasiparticle random phase approximation, Kamlah expansion
\cite{[Kam68],[Zhe92]},
Lipkin-Nogami (LN) method
\cite{[Lip60],[Nog64],[Gal94],[Rei96],[Cwi96],[Val97],[Val00a],[Ben00e],[Sto03]},
the particle-number  projection after variation (PNPAV) \cite{[RS80],[Ang02]},
the projected LN method (PLN)
\cite{[Dob93],[Mag93],[Ang02],[Sto03],[Sam04]}, and
the variation after particle-number projection ({\VAP})
\cite{[Die64],[Sch87],[She00a],[Ang01a],[Ang02]}.

In this work, we concentrate on
the {\VAP} method.
Recently,  it has been shown \cite{[She00a]}
that the total energy in the
HFB+{\VAP}  approach can be expressed as a
functional of the unprojected HFB density matrix and pairing tensor.
Its variation leads to a set of HFB-like equations with modified
self-consistent  fields. The method has been
illustrated within schematic models \cite{[She02]}, and also
implemented in the HFB calculations with  the finite-range Gogny force
\cite{[Ang01a],[Ang02]}.
Here, we adopt it for the Skyrme energy-density
functionals and zero-range  pairing forces; in this case the building
blocks of the method  are  the local particle-hole and particle-particle
 densities and mean fields.
In the present study, the HFB equations are solved by using the Harmonic Oscillator
(HO) basis, but the formalism can be straightforwardly applied with the
Transformed Harmonic Oscillator (THO) basis \cite{[Sto03],[Sto05]},
which helps maintain the correct asymptotic behavior of the
single-quasiparticle wave functions.

 The paper is organized as follows.
 Section~\ref{sec1} gives a
brief overview of the HFB theory and defines the densities and fields
entering the formalism.
 Section~\ref{sec2} extends the {\VAP}
method of Ref.~\cite{[She02]} to the case of
the HFB theory with Skyrme interaction.
 The technical details pertaining  to the Skyrme HFB+{\VAP} method
 are given in
 Sec.~\ref{sec3}, while Sec.~\ref{sec5} contains an
illustrative example of calculations for the Ca  and Sn isotopes.
In particular, the   LN and PLN  approximations are
compared to the {\VAP} results.
Summary and discussion are given  in Section~\ref{sec6}. Preliminary
results of our {\VAP} calculations were presented in Ref.~\cite{[Sto05b]}.

\section{The HFB method}
\label{sec1}

The many-body Hamiltonian of a system of fermions is usually expressed in
terms of a set of annihilation and creation operators $(c,c^{\dagger })$:
\begin{eqnarray}
H &=&\sum_{nn'}e_{nn'}~c_{n}^{\dagger }c_{n'}  \nonumber
\\
&+&\tfrac{1}{4}\sum_{nn'mm'}V_{nn'mm'}~c_{n}^{\dagger }c_{n'}^{\dagger
}c_{m'}c_{m},  \label{eq:13}
\end{eqnarray}
where
\begin{equation}
V_{nn'mm'}=\langle nn'|V|mm'-m'm\rangle  \label{eq:23}
\end{equation}
are the anti-symmetrized two-body interaction matrix-elements.

In the HFB method, the ground-state wave function is the
quasiparticle vacuum $|\Phi \rangle ,$ defined as $\alpha
_{k}|\Phi \rangle =0$, where the quasiparticle operators $(\alpha
,\alpha^{\dagger })$ are connected to the original particle
operators via  the Bogoliubov  transformation
\begin{eqnarray}
\quad \quad \alpha_{k} &=&\sum_{n}\left( U_{nk}^{\ast
}c_{n}+V_{nk}^{\ast
}c_{n}^{\dagger }\right) ,  \label{eq:43} \\
\quad \quad \alpha_{k}^{\dagger } &=&\sum_{n}\left(
V_{nk}c_{n}+U_{nk}c_{n}^{\dagger }\right),  \label{eq:53}
\end{eqnarray}
where the matrices $U$ and $V$ satisfy the unitarity and completeness relations:
\begin{eqnarray}
U^{\dagger }U+V^{\dagger }V=I,&&UU^{\dagger
}+V^{\ast }V^{T}=I,  \label{eq:83} \\
U^{T}V+V^{T}U=0, &&UV^{\dagger }+V^{\ast }U^{T}=0. \label{eq:93}
\end{eqnarray}

\subsection{The HFB equations}

In terms of the density matrix $\rho $ and pairing tensor $\kappa $,
defined as
\begin{equation}
\rho =V^{\ast }V^{T},\quad \kappa =V^{\ast }U^{T}=-UV^{\dagger },
\label{eq:132}
\end{equation}
the HFB energy is expressed as an energy functional:
\begin{eqnarray}
E[\rho ,\kappa ] &=&\frac{\langle \Phi |H|\Phi \rangle }{\langle
\Phi
|\Phi \rangle }  \nonumber \\
&=&{\rm Tr}\left[ \left(e+\tfrac{1}{2}\Gamma \right)\rho \right]
-\tfrac{1}{2}{\rm Tr} \left[ \Delta \kappa^{\ast }\right] ,
\label{ehfb}
\end{eqnarray}
where
\begin{eqnarray}
\Gamma_{nm}
&=&\sum_{n'm'}V_{nn'mm'}
\rho_{m'n'},  \label{eq:1302} \\
\Delta_{nn'} &=&\tfrac{1}{2}\sum_{mm'}V
_{nn'mm'}\kappa_{mm'}.  \label{eq:1303}
\end{eqnarray}
The variation of the HFB energy (\ref{ehfb}) with respect to $\rho$ and $\kappa$
yields  the HFB equations:
\begin{equation}  \label{hfb}
{\cal H}\left(
\begin{array}{c}
U_k \\
V_k
\end{array}
\right) =E_k\left(
\begin{array}{c}
U_k \\
V_k
\end{array}
\right) ,
\end{equation}
where
\begin{equation}
{\cal H}=\left(
\begin{array}{cc}
e+\Gamma -\lambda & \Delta \\
-\Delta^{\ast } & -(e+\Gamma )^{\ast }+\lambda
\end{array}
\right) ,  \label{hfbeq}
\end{equation}
$U_k$ and $V_k$ are the $k$th columns of matrices $U$ and $V$,
respectively, and $E_k$ is a positive quasiparticle energy eigenvalue.
Since the HFB state  $|\Phi\rangle$ violates the particle-number symmetry,
the Fermi energy  $\lambda $ is introduced to  fix the average particle number.

\subsection{The Skyrme HFB method}

For the zero-range  Skyrme forces, the HFB formalism can be written
directly in the coordinate representation \cite{[Bul80],[Dob84],[Per04]} by introducing
particle and pairing densities
\begin{eqnarray}
\rho ({\bm r}\sigma ,{\bm r^{\prime }}\sigma^{\prime })
&=&\tfrac{1}{2} \rho ({\bm r},{\bm r}^{\prime })\delta_{\sigma
\sigma^{\prime }}\nonumber \\ &+& \tfrac{1}{2} \sum_{i}(\sigma
|\sigma_{i}|\sigma^{\prime })\rho_{i}({\bm r},{\bm r}
^{\prime }) ,  \label{densitieemd} \\
\tilde{\rho}({\bm r}\sigma ,{\bm r^{\prime }}\sigma^{\prime })
&=&\tfrac{1}{2 } \tilde{\rho}({\bm r},{\bm r}^{\prime })\delta
_{\sigma \sigma^{\prime }}\nonumber \\ &+&\tfrac{1}{2}
\sum_{i}(\sigma |\sigma_{i}|\sigma^{\prime })\tilde{\rho}_{i}( {\bm
r},{\bm r}^{\prime }),  \label{densitieemp}
\end{eqnarray}
which explicitly depend on spin.
The use of the pairing density  $\tilde{\rho}$,
\begin{equation}
\tilde{\rho}({\bm r}\sigma ,{\bm r^{\prime }}\sigma^{\prime
})=-2\sigma^{\prime }\kappa ({\bm r,}\sigma ,{\bm r^{\prime
},}-\sigma^{\prime }),
\end{equation}
instead of the pairing tensor $\kappa $ is convenient when
restricting to time-even quasiparticle states where both $\rho $
and $\tilde{\rho}$ are hermitian and time-even \cite{[Dob84]}.

The densities $\rho$ and  $\tilde\rho$ can be expressed in the
single-particle basis:
\begin{eqnarray}
\rho ({\bm r}\sigma ,{\bm r^{\prime }}\sigma^{\prime })
&=&\sum_{nn^{\prime }}\rho_{nn^{\prime }}~\psi_{n^{\prime }}^{\ast
}({\bm r^{\prime }},\sigma
^{\prime })\psi_{n}({\bm r},\sigma ),  \label{denmc} \\
\tilde{\rho}({\bm r}\sigma ,{\bm r^{\prime }}\sigma^{\prime })
&=&\sum_{nn^{\prime }}\tilde{\rho}_{nn^{\prime }}~\psi_{n^{\prime
}}^{\ast }({\bm r^{\prime }},\sigma^{\prime })\psi_{n}({\bm
r},\sigma ), \label{pdenmc}
\end{eqnarray}
where $\rho_{n^{\prime }n}$ and  $\tilde\rho_{n^{\prime }n}$ are
the corresponding density matrices. In this study,
we take  ${\psi_{n}({\bm r},\sigma )}$ as a set of the
HO wave functions.

The building blocks of the Skyrme HFB method are the local densities, namely
the  particle  density $\rho ({\bm r})$, kinetic energy
density $ \tau ({\bm r})$, and spin-current density ${\mathsf
J}_{ij}({\bm r})$:
\begin{equation}
\begin{array}{ccl}
\rho ({\bm r}) & = & \rho ({\bm r},{\bm r}), \\
~ &  &  \\
\tau ({\bm r}) & = & \left. \nabla_{{\bm r}}\nabla_{{\bm r}^{\prime
}}\rho
({\bm r},{\bm r}^{\prime })\right|_{{\bm r}^{\prime }={\bm r}}\;, \\
~ &  &  \\
{\mathsf J}_{ij}({\bm r}) & = & \tfrac{1}{2i}\left. \left( \nabla
_{i}-\nabla_{i}^{\prime }\right) \rho_{j}({\bm r},{\bm r}^{\prime
})\right|_{{\bm r}^{\prime }={\bm r}}\;,
\end{array}
\label{densities}
\end{equation}
as well as the corresponding  pairing  densities
 $\tilde{\rho}({\bm r})$,  $\tilde{\tau}({\bm r})$ and
$\tilde{\mathsf J}_{ij}({\bm r})$.

In the coordinate representation, the Skyrme HFB
energy (\ref{ehfb})  can be written as a
functional of the local particle and pairing densities:
\begin{equation}
E[\rho,\tilde{\rho}]=\frac{\langle \Phi |H|\Phi \rangle }{\langle
\Phi |\Phi \rangle } =\int d{\bm r}~{\cal H}({\bm r}). \label{shfb}
\end{equation}
The energy density $ {\cal
H}({\bm r})$  is a sum
of the particle  $H({\bm r})$ and pairing energy density
$\tilde{H}({\bm r})$:
\begin{equation}
{\cal H}({\bm r})=H({\bm r})+\tilde{H}({\bm r}).  \label{skyrmeh}
\end{equation}
The  derivatives of $E[\rho ,\tilde{\rho}]$ with respect to density
matrices $\rho$ and $\tilde{\rho}$
define the self-consistent particle
($h$)   and pairing
($\tilde{h}$) fields, respectively.
The explicit expressions for $H({\bm r})$,
$\tilde{H}({\bm r})$, $h({\bm r},\sigma ,\sigma^{\prime })$, and
$\tilde{h}({\bm r} ,\sigma ,\sigma^{\prime })$ have been given
in Ref.~\cite{[Dob84]} and will not be repeated here.

The Skyrme HFB equations can be written in the matrix form as:
\begin{equation}
\left(
\begin{array}{cc}
h-\lambda & \tilde{h} \\
\tilde{h} & -h+\lambda
\end{array}
\right) \left(
\begin{array}{c}
\varphi_{1,k} \\
\varphi_{2,k}
\end{array}
\right) =E_k \left(
\begin{array}{c}
\varphi_{1,k} \\
\varphi_{2,k}
\end{array}
\right), \label{shfbeq}
\end{equation}
where
\begin{eqnarray}
h_{nn^{\prime }} &=&\frac{\partial E[\rho ,\tilde{\rho}]}{\partial
\rho
_{n^{\prime }n}}  \nonumber \\
&=&\sum_{\sigma \sigma^{\prime }}\int d{\bm r}~\psi_{n}^{\ast }({\bm
r} ,\sigma )h({\bm r},\sigma ,\sigma^{\prime })\psi_{n^{\prime
}}({\bm r} ,\sigma^{\prime }),  \label{ph}
\end{eqnarray}
and
\begin{eqnarray}
\tilde{h}_{nn^{\prime }} &=&\frac{\partial E[\rho ,\tilde{\rho}]}{\partial
\tilde{\rho}_{n^{\prime }n}}  \nonumber \\
&=&\sum_{\sigma \sigma^{\prime }}\int d{\bm r}~\psi_{n}^{\ast }( {\bm r},\sigma )\tilde{h}({\bm r},\sigma ,\sigma
^{\prime })\psi_{n^{\prime}}({\bm r} ,\sigma^{\prime }), \label{pp}
\end{eqnarray}
and $\varphi_{1,k}$ and $\varphi_{2,k}$ are the upper and lower
components, respectively, of the quasiparticle wave function corresponding to
the positive quasiparticle energy $E_k$. After solving the HFB equations
(\ref{shfbeq}), one obtains the density matrices,
\begin{eqnarray}
\rho_{nn^{\prime }} = \sum_{E_k>0} \varphi_{2,nk} \varphi^*_{2,n^{\prime }k} , \label{denmc2} \\
\tilde{\rho}_{nn^{\prime }} = -\sum_{E_k>0} \varphi_{2,nk} \varphi^*_{1,n^{\prime }k} ,  \label{pdenmc2}
\end{eqnarray}
which define the spatial densities (\ref{denmc}) and (\ref{pdenmc}) .

We note in passing that the derivation of the coordinate-space
HFB equations
\cite{[Dob84]} is strictly
valid only when the time-reversal symmetry is assumed.
When the time-reversal
symmetry is broken, one has to introduce additional  real
vector particle densities ${\bm s}$, ${\bm j}$,
${\bm T}$ \cite{[Eng75]}, while the pairing densities acquire
imaginary parts; see Ref.~\cite{[Per04]} for complete derivations.

\section{Variation after particle-number projection}
\label{sec2}

\subsection{The HFB+{\VAP} method}\label{VAPgen}

It has been demonstrated \cite{[She00a]} that the
HFB+{\VAP}   energy,
\begin{eqnarray}
E^{N}[\rho ,\kappa]&=&\frac{\left\langle \Phi |HP^{N}|\Phi \right\rangle }{
\left\langle \Phi |P^{N}|\Phi \right\rangle }  \nonumber \\
&=&\frac{\int d\phi \langle \Phi |{H}e^{i\phi ({\hat{N}}-N)}|\Phi
\rangle }{ \int d\phi \langle \Phi |e^{i\phi ({\hat{N}}-N)}|\Phi
\rangle },  \label{eqpnp}
\end{eqnarray}
where $P^{N}$ is the particle-number projection operator,
\begin{equation}
P^{N}=\frac{1}{2\pi }\int d\phi \ e^{i\phi (\hat{N}-N)},  \label{eq:232}
\end{equation}
can be written as an energy functional of the unprojected densities
(\ref{eq:132}).

The variation of Eq.~(\ref{eqpnp})  results in
the HFB+{\VAP}  equations:
\begin{equation}
{\cal H}^{N}\left(
\begin{array}{c}
U_k^N \\
V_k^N
\end{array}
\right) ={\cal E}_k\left(
\begin{array}{c}
U_k^N \\
V_k^N
\end{array}
\right) ,  \label{phfb}
\end{equation}
where
\begin{equation}
{\cal H}^{N}=\left(
\begin{array}{cc}
\varepsilon^{N}+\Gamma^{N}+\Lambda^{N} & \Delta^{N} \\
-{(\Delta^{N})}^{\ast } & -(\varepsilon^{N}+\Gamma^{N}+\Lambda
^{N})^{\ast }
\end{array}
\right) .  \label{phfbeq}
\end{equation}
Equations~(\ref{phfb}) and (\ref{phfbeq}) have the same structure as
Eqs.~(\ref{hfb}) and (\ref{hfbeq}), except that the expressions for
the {\VAP} fields are now different \cite{[She00a],[She02]}, i.e.,
\begin{eqnarray}
\varepsilon^{N} &=&\tfrac{1}{2}\int d\phi \,\,y(\phi )\left\{ Y(\phi
){\rm Tr}
[e\rho (\phi )]\right.  \nonumber \\
&+&\left. [1-2ie^{-i\phi }\sin \phi \rho (\phi )]eC(\phi )\right\}
+{\rm h.c.}, \label{eN-pnp}
 \\ \nonumber \\
\Gamma^N &=& \tfrac{1}{4}\int d\phi \,\,y(\phi ) \left\{Y(\phi) {\rm
Tr}
[\Gamma(\phi)\rho(\phi)] \right.  \nonumber \\
&+& \left. 2[1-2ie^{-i\phi}\sin\phi\rho(\phi)]\Gamma(\phi)C(\phi)\right\} +
{\rm h.c.},  \label{GN-pnp}
 \\ \nonumber \\
\Delta^{N}&=&\tfrac{1}{2}\int d\phi \;y(\phi )e^{-2i\phi }C\left(
\phi \right) \Delta (\phi )-(...)^{T},  \label{D-pnp}
 \\ \nonumber \\
\Lambda^{N} &=&-\tfrac{1}{4}\int d\phi \,\,y(\phi )\left\{ Y(\phi
){\rm Tr}
[\Delta (\phi )\overline{\kappa }^{\ast }(\phi )]\right.  \nonumber \\
&-&\left. 4ie^{-i\phi }\sin \phi \;C(\phi )\Delta (\phi
)\overline{\kappa }^{\ast }(\phi )\right\} + {\rm h.c.},  \label{LN-pnp}
\end{eqnarray}
with
\begin{eqnarray}
\Gamma_{nm}(\phi ) &=&\sum_{n'm'}V
_{nn'mm'}\rho_{m'n'}(\phi ),  \label{Gmatr} \\
\Delta_{nn'}(\phi )
&=&\tfrac{1}{2}\sum_{mm'}V_{nn'mm'}\kappa_{mm'}(\phi ),  \label{Dmatr}
\end{eqnarray}
where, using the unit matrix $\hat{I}$,
\begin{eqnarray}
\rho (\phi ) &=&C(\phi )\rho ,  \label{rphi} \\
\kappa (\phi ) &=&C(\phi )\kappa,  \label{kphi} \\
\overline{\kappa }(\phi ) &=&e^{2i\phi
}C^{\dagger }(\phi )\kappa ,  \label{kbhi} \\
C(\phi ) &=&e^{2i\phi }\left[ 1+\rho (e^{2i\phi }-1)\right]^{-1},
\label{cphi} \\
x(\phi ) &=&\frac{1}{2\pi }\frac{e^{-i\phi N}\det (e^{i\phi
}I)}{\sqrt{
\det C(\phi )}},  \label{xphi} \\
y(\phi ) &=&\frac{x(\phi )}{\int d\phi^{\prime }\,x(\phi^{\prime
})},  \label{yphi}
\end{eqnarray}
and
\begin{eqnarray}
Y(\phi ) &=&\,ie^{-i\phi }\sin \phi \;C(\phi )  \nonumber \\
&-&i\int d\phi^{\prime }y(\phi^{\prime })e^{-i\phi^{\prime }}\sin
\phi^{\prime }\;C(\phi^{\prime }).  \label{ynm}
\end{eqnarray}
After solving the HFB+{\VAP} equations (\ref{phfb}), one obtains the
intrinsic density matrix and pairing tensor:
\begin{equation}
\rho =(V^N)^{\ast }(V^N)^{T},\quad \kappa =(V^N)^{\ast }(U^N)^{T}.
\label{intrden}
\end{equation}

Finally, the total  HFB+{\VAP}  energy is given by
\begin{eqnarray}
E^N[\rho ,\kappa] &=&\int d\phi ~y(\phi )~{\rm Tr}\left( e\rho (\phi )+\tfrac{1}{2}
\Gamma (\phi )\rho (\phi )\right)  \nonumber \\
&-&\int d\phi ~y(\phi )~\tfrac{1}{2}{\rm Tr}\left( \Delta (\phi
)\overline{ \kappa }^{\ast }(\phi )\right) .  \label{EPNPP}
\end{eqnarray}
The quantity $y(\phi )$ plays a role of an $N$-dependent metric. The
integrands in Eqs.~(\ref{eN-pnp})--(\ref{D-pnp}) take the familiar  HFB
limit at $\phi$=0, while the integrand in (\ref{LN-pnp}) vanishes
($\Lambda^{N}$ does not appear in the standard HFB approach).

\subsection{The Lipkin-Nogami method}\label{LNgen}

The LN method \cite{[Lip60],[Nog64]} constitutes an astute
and efficient way of performing an {\em approximate} {\VAP} calculation. It can be
considered \cite{[Flo97]}
as a variant of the second-order Kamlah expansion
\cite{[Kam68],[Zhe92]}, in which the {\VAP} energy (\ref{eqpnp})
is approximated by a simple expression,
\begin{equation}
E_{\text{LN}}=E[\rho ,\tilde{\rho}] - \lambda_2(\langle\hat{N}^2\rangle-N^2),
\label{enekam}
\end{equation}
with $\lambda_2$ depending on the HFB state $|\Phi\rangle$ and
representing the curvature of the {\VAP} energy with respect to the
particle number. The role of $\lambda_2$ in the Kamlah and LN methods
differs. In the former, $\lambda_2$ is varied along with variations of the
HFB state  $|\Phi\rangle$, while in the latter, this variation is
neglected. Had the second-order Kamlah expression (\ref{enekam}) been
exact, the variation of $\lambda_2$ would have been fully justified
and the method would be giving the exact {\VAP} energy. However,
since the second-order expression is, in practical applications,
never exact, it is usually more reasonable to adopt the LN
philosophy, in which one rather strives to find the best estimate of
the curvature $\lambda_2$ instead of finding it variationally in an
approximate way.

When the HFB method is applied to a given Hamiltonian, values of
$\lambda_2$ can be estimated by calculating new mean-field
potentials, $\Gamma'$ and $\Delta'$, that are analogous to the
standard mean fields of Eqs.~(\ref{eq:1302}) and (\ref{eq:1303});
see, e.g., Refs.~\cite{[Flo97],[Sto03]}. However, apart from studies
based on the Gogny Hamiltonian \cite{[Ang02]}, such a formula was not
used, because most often the self-consistent calculations are
performed within the density functional approach or by using
different interactions in the particle-hole and particle-particle
channels. Moreover, in most studies, such as those of
 Ref.~\cite{[Gal94]}, the terms in $\lambda_2$
originating from the particle-hole channel are simply
disregarded.

Similarly, as in our previous study \cite{[Sto03]}, here we adopt
an efficient phenomenological way of estimating the curvature $\lambda_{2}$
from the seniority-pairing expression,
\begin{equation}\label{ll3}
\lambda_{2}=\frac{G_{\text{eff}}}{4} \frac {{\rm Tr'} (1-\rho)\kappa~ {\rm Tr'} \rho \kappa  - 2~{\rm Tr} (1-\rho)^2
\rho^2} {\left[{\rm Tr}\rho (1-\rho )\right]^{2}-2~{\rm Tr}\rho^{2}(1-\rho)^{2}},
\end{equation}
where the effective pairing strength,
\begin{equation}\label{ll4}
G_{\text{eff}} = -\frac{\bar{\Delta}^2}{E_{\text{pair}}},
\end{equation}
is determined from the HFB pairing energy,
\begin{equation}\label{ll5}
E_{\text{pair}} = -\frac{1}{2}{\rm Tr}\Delta \kappa^* ,
\end{equation}
and the average pairing gap \footnote{
Along with the standard trace of a matrix $A$, ${\rm Tr}A$=$\sum_n A_{nn}$,
in Eqs.~(\ref{ll3})--(\ref{ll6}) we use ${\rm Tr'}A$=$\sum_n A_{n\bar{n}}$.},
\begin{equation}\label{ll6}
\bar{\Delta}  = \frac{{\rm Tr'}\Delta \rho}{{\rm Tr}\rho} \, .
\end{equation}
Expression (\ref{ll3}) pertains to a system of particles occupying
single-particle levels with fixed (non-self-consistent) energies and
interacting with a seniority pairing interaction. In our method, this
expression is used to probe the density of self-consistent energies
that determine the curvature $\lambda_{2}$. All quantities defining
$\lambda_{2}$ in Eq.~(\ref{ll3})
depend on the self-consistent solution and
microscopic interaction, while the effective pairing strength
$G_{\text{eff}}$ is only an auxiliary quantity. The quality of the
prescription for calculating $\lambda_{2}$ can be tested against the
exact {\VAP} results (see Sec.~\ref{sec5}).

\subsection{The Skyrme HFB+{\VAP} method}

Following the {\VAP} procedure of Sec.~\ref{VAPgen},
 one can  develop the Skyrme HFB+{\VAP} equations
by introducing the gauge-angle-dependent transition
density matrices:
\begin{eqnarray}
\rho ({\bm r}\sigma ,{\bm r^{\prime }}\sigma^{\prime },\phi )
&=&\sum_{nn^{\prime }}\rho_{nn^{\prime }}(\phi )~\psi_{n^{\prime
}}^{\ast }({\bm r^{\prime }},\sigma^{\prime })\psi_{n}({\bm
r},\sigma ),
\label{denmcp} \\
\tilde{\rho}({\bm r}\sigma ,{\bm r^{\prime }}\sigma^{\prime },\phi
) &=&\sum_{nn^{\prime }}\tilde{\rho}_{nn^{\prime }}(\phi )
~\psi_{n^{\prime }}^{\ast }({\bm r^{\prime }},\sigma^{\prime })
\psi_{n}({\bm r},\sigma ). \label{penmcp}
\end{eqnarray}
In the above equation, the density matrix $\rho_{nn^{\prime }}(\phi )$
is given by Eq.~(\ref{rphi}) while
\begin{equation}\label{prphis}
\tilde{\rho}(\phi ) =e^{-i\phi }C(\phi )\tilde{\rho}.
\end{equation}
The associated gauge-angle-dependent local densities  $\rho ({\bm
r},\phi )$, $\tau ({\bm r},\phi )$, ${\mathsf J}_{ij}({\bm r },\phi )$, $
\tilde{\rho}({\bm r},\phi )$, $\tilde{\tau}({\bm r},\phi )$, and
$\tilde{\mathsf J}_{ij}({\bm r},\phi )$ are defined by
Eqs.~(\ref{densities}) in terms of  the density matrices
(\ref{denmcp}) and (\ref{penmcp}). Using  the Wick
theorem for matrix elements \cite{[RS80]}, one can show that the gauge-angle-dependent
transition energy density  ${\cal H}({\bm r},\phi )$ can be obtained  from the
intrinsic  energy density $ {\cal H}({\bm r})$ simply by
substituting particle (pairing) local densities with their
gauge-angle-dependent counterparts (e.g., $\rho ({\bm
r})$$\rightarrow$$\rho ({\bm r},\phi )$).

In the case of Skyrme functionals, the HFB+{\VAP} energy (\ref{eqpnp})
can be expressed through  an integral
\begin{equation}
E^{N}[\rho ,\tilde{\rho}]=\int d\phi ~y(\phi )~E(\phi ), \label{shfbN}
\end{equation}
where the transition energy reads:
\begin{equation}
E(\phi )=\frac{\langle \Phi |He^{i\phi \hat{N}}|\Phi \rangle }{\langle \Phi
|e^{i\phi \hat{N}}|\Phi \rangle }=\int d{\bm r}~{\cal H}({\bm r},\phi ).
\label{shfbphi}
\end{equation}
The projected energy (\ref{shfbN})
is  a functional $E^{N}[\rho ,\tilde{\rho}]$ of the matrix
elements of intrinsic (i.e., $\phi$=0) matrices $\rho$ and $\tilde{\rho}$.

In order to compute  the derivatives of $E^{N}(\rho ,\tilde{\rho})$
with respect to $\rho $ and $\tilde{\rho}$, one should take first
the derivatives of $ E^{N}[\rho ,\tilde{\rho}]$ with respect to
$\rho (\phi )$ and $\tilde{\rho} (\phi )$,
and then the derivatives of $\rho (\phi
)$ and $\tilde{\rho}(\phi )$ with respect to the intrinsic densities $\rho $
and $\tilde{\rho}$.
For example,
\begin{eqnarray}
\frac{\partial E^{N}[\rho ,\tilde{\rho}]}{\partial \rho
_{nn^{\prime }}} &=&\int d\phi ~y(\phi )~\left[ {\frac{1}{y(\phi
)}}\frac{\partial y(\phi )}{
\partial \rho_{nn^{\prime }}}~E(\phi )\right.  \nonumber \\
&+&\sum_{\alpha \beta }\frac{\partial E(\phi )}{\partial
\rho_{\alpha \beta }(\phi )}\frac{\partial \rho_{\alpha \beta
}(\phi )}{\partial \rho
_{nn^{\prime }}}  \nonumber \\
&+&\sum_{\alpha \beta }\frac{\partial E(\phi )}{\partial
\tilde{\rho}_{\alpha \beta }(\phi )}\frac{\partial
\tilde{\rho}_{\alpha \beta }(\phi )}{
\partial \tilde{\rho}_{nn^{\prime }}}  \nonumber \\
&+&\left. \sum_{\alpha \beta }\frac{\partial E(\phi )}{\partial
\tilde{\rho}_{\alpha \beta }^{\ast }(-\phi )}\frac{\partial
\tilde{\rho}_{\alpha \beta }^{\ast }(-\phi )}{\partial \rho
_{nn^{\prime }}}\right] .  \label{examp1}
\end{eqnarray}
With the use of the identity:
\begin{equation}
\delta_{mm^{\prime }}-2ie^{-i\phi }\sin \phi ~\rho_{mm^{\prime
}}(\phi )=e^{-2i\phi }C_{mm^{\prime }}(\phi ),
\end{equation}
the partial derivatives in Eq.~(\ref{examp1}) can easily be
calculated:
\begin{eqnarray}
\frac{\partial y(\phi )}{\partial \rho_{nn^{\prime }}}
&= & y(\phi
)~Y_{nn^{\prime }}(\phi ),  \label{der1} \\
\frac{\partial \rho_{mm^{\prime }}(\phi )}{\partial \rho
_{nn^{\prime }}}
&=&\delta_{m^{\prime }n^{\prime }}C_{mn}(\phi )  \nonumber \\
&-&2ie^{-i\phi }\sin(\phi )\rho_{n^{\prime }m^{\prime }}(\phi
)C_{mn}(\phi ),
\\
\frac{\partial \tilde{\rho}_{mm^{\prime }}(\phi )}{\partial \rho
_{nn^{\prime }}} &=&-2ie^{-i\phi }\sin(\phi )\tilde{\rho}_{n^{\prime
}m^{\prime }}(\phi )C_{mn}(\phi ), \\
\frac{\partial \tilde{\rho}_{mm^{\prime }}(\phi )}{\partial
\tilde{\rho}_{nn^{\prime }}} &=&e^{-i\phi }C_{mn}(\phi )\delta
_{m^{\prime }n^{\prime }}
\nonumber \\
&+&e^{-i\phi }C_{m\bar{n}}(\phi )\delta_{\bar{n}m^{\prime
}}s_{\bar{n}^{\prime }}s_{\bar{n}}^{\ast },  \label{der2}
\end{eqnarray}
where $\bar{n}$ and $ s_{n}$ ($s_{n}s_{n}^{\ast }=1$,
$s_{\bar{n}}=-s_{n}$) are defined using the time-reversal operator
$\hat{T}$, as
\begin{equation}
\hat{T}\psi_{n}({\bm r},\sigma )=s_{n}\psi_{\bar{n}}({\bm r},\sigma
).
\end{equation}

By  inserting Eqs.~(\ref{der1})--(\ref{der2}) in
Eq.~(\ref{examp1}), the latter reads
\begin{eqnarray}
&&\frac{\partial E^{N}[\rho ,\tilde{\rho}]}{\partial \rho } =\int d\phi
~y(\phi )~Y(\phi )~E(\phi )  \nonumber \\
&&+\int d\phi ~y(\phi )~e^{-2i\phi }~C(\phi )h(\phi )C(\phi )  \\
&&-\left[ \int d\phi ~y(\phi )~2ie^{-i\phi }\sin(\phi
)~\tilde{\rho}(\phi )
\tilde{h}(\phi )C(\phi ) + {\rm h.c.}\right],  \nonumber
\end{eqnarray}
where
\begin{eqnarray}
&&h_{nn^{\prime }}(\phi ) = \frac{\partial E(\phi )}{\partial \rho_{n^{\prime }n} (\phi )} \nonumber \\
&&= \sum_{\sigma \sigma^{\prime }}\int
d^{3}{\bm r} ~\psi_{n}^{\ast }({\bm r}\sigma )h({\bm r,}\sigma
,\sigma^{\prime },\phi )\psi_{n^{\prime }}({\bm r}\sigma^{\prime
}),  \label{phphi}  \\
&&\tilde{h}_{nn^{\prime }}(\phi ) =
\frac{\partial E(\phi )}{\partial \tilde{\rho}_{n^{\prime }n}(\phi )} \nonumber \\
&&= \sum_{\sigma \sigma^{\prime }}\int
d^{3} {\bm r}~\psi_{n}^{\ast }({\bm r}\sigma
)\tilde{h}({\bm r},\sigma ,\sigma^{\prime },\phi )\psi_{n^{\prime }}({\bm
r}\sigma^{\prime }).  \label{ppphi}
\end{eqnarray}
The derivative of $ E^{N}(\rho ,\tilde{\rho})$ with respect to
 $\tilde{\rho}$ can be computed in a similar manner.
The $\phi$-dependent fields $h({\bm r},\sigma ,\sigma
^{\prime },\phi )$ and $\tilde{h}({\bm r},\sigma ,\sigma
^{\prime },\phi )$ are obtained  by substituting the
local particle and pairing densities  in the
intrinsic fields $h({\bm r},\sigma ,\sigma^{\prime })$  and
$\tilde{h}({\bm r},\sigma ,\sigma^{\prime })$ with their
gauge-angle-dependent counterparts.

The Skyrme HFB+{\VAP} equations can finally be written
in the form
\begin{equation}
\left(
\begin{array}{cc}
h^{N} & \tilde{h}^{N} \\
\tilde{h}^{N} & -h^{N}
\end{array}
\right) \left(
\begin{array}{c}
\varphi^{N}_{1,k} \\
\varphi^{N}_{2,k}
\end{array}
\right) ={\cal E}_k\left(
\begin{array}{c}
\varphi^{N}_{1,k} \\
\varphi^{N}_{2,k}
\end{array}
\right) ,  \label{pshfbeq}
\end{equation}
with particle--hole and particle--particle Hamiltonians
\begin{eqnarray}
&&h^{N} =\int d\phi y(\phi )Y(\phi )E(\phi )  \nonumber \\
&&+\int d\phi y(\phi )e^{-2i\phi }~C(\phi )h(\phi )C(\phi ) \label{hph} \\
&&-\left[  \int d\phi y(\phi )2ie^{-i\phi }\sin(\phi
)\tilde{\rho}(\phi )
\tilde{h}(\phi )C(\phi ) +  {\rm h.c.}\right],   \nonumber \\
\tilde{h}^{N} &=&\int d\phi y(\phi )e^{-i\phi }\left[
\tilde{h}(\phi )C(\phi )+(...)^{T}\right]. \label{hpp}
\end{eqnarray}
Finally, solutions of the HFB+{\VAP} equations
(\ref{pshfbeq}) allow for calculating the intrinsic density matrices as,
\begin{eqnarray}
\rho_{nn^{\prime }} = \sum_{E_k>0} \varphi^N_{2,nk} \varphi^{N*}_{2,n^{\prime }k} , \label{denmc3} \\
\tilde{\rho}_{nn^{\prime }} = -\sum_{E_k>0} \varphi^N_{2,nk} \varphi^{N*}_{1,n^{\prime }k} .  \label{pdenmc3}
\end{eqnarray}

Let us re-emphasize that the densities and fields that enter
the Skyrme HFB+{\VAP} equations
are immediate generalizations of  the analogous quantities that
appear in the standard
Skyrme HFB formalism. Of course, due to
   the presence of  $C(\phi)$ and  integrations over the gauge angle,
the Skyrme HFB+{\VAP}
calculations are appreciably  more involved.

\section{Skyrme HFB+{\VAP} procedure: practical details}
\label{sec3}

\subsection{Two kinds of nucleons}

As one is dealing with  $Z$ protons and $N$ neutrons, two gauge angles,
$\phi_{n}$ and $\phi_{p}$, must enter
the number projection operator:
\begin{equation}
P^{NZ}=\frac{1}{2\pi }\int d\phi_{n}\ e^{i\phi
_{n}(\hat{N}-N)}\frac{1}{ 2\pi }\int d\phi_{p}\ e^{i\phi
_{p}(\hat{Z}-Z)}.  \label{eq:232NZ}
\end{equation}
Consequently, the total projected  energy (\ref{shfbN}) becomes
a double integral,
\begin{equation}
E^{N}=\int d\phi_{n}~d\phi_{p}~y_{n}(\phi_{n})~y_{p}(\phi_{p})~
E(\phi_{n},\phi_{p}), \label{enen}
\end{equation}
where the transition energy density
\begin{equation}
E(\phi_{n},\phi_{p})=\int d{\bm r}~{\cal H}({\bm r},\phi_{n},\phi
_{p})  \label{shfbphi2}
\end{equation}
depends on both gauge angles $\phi_{n}$, $\phi_{p}$.

To simplify notation, we   use the isospin
 label $q$=$\tau_3$ ($q$=+1 for neutrons and --1 for protons)
 and $\bar{q}$=$-q$. In the following, we shall employ the convention
 $y(\phi_{q})$$\equiv$$y_q(\phi_{q})$, $C(\phi_{q})$$\equiv$$C_q(\phi_{q})$,
 and $Y(\phi_{q})$$\equiv$$Y_q(\phi_{q})$.
The isospin-dependent particle-hole and
particle-particle fields (\ref{hph}), (\ref{hpp}) can be written as:
\begin{eqnarray}
h_{q}^{N} &=&\int d\phi_{q}~y(\phi_{q})~  \nonumber \\
&\times &\left[ Y(\phi_{q})\left( \int y(\phi_{\bar{q}})~E(\phi
_{q},\phi_{\bar{q}})d\phi_{\bar{q}}~\right) \right.  \nonumber \\
&+&\left. e^{-2i\phi_{q}}~C(\phi_{q})\left( \int y(\phi_{\bar{q
}})~h_{q}(\phi_{q},\phi_{\bar{q}})d\phi_{\bar{q}}~\right) C(\phi
_{q})
\right]  \nonumber \\
&-&\left[ \int d\phi_{q}~y(\phi_{q})\right. 2ie^{-i\phi
_{q}}\sin(\phi
_{q})\tilde{\rho}_q(\phi_{q})  \nonumber \\
&\times &\left. \left( \int y(\phi_{\bar{q}})\tilde{h}_{q}(\phi
_{q},\phi_{\bar{q}})d\phi_{\bar{q}}\right) C(\phi
_{q})+^{\;}h.c.\right] ,
\label{hphtotph} \\
\tilde{h}_{q}^{N} &=&\int d\phi_{q}~y(\phi_{q})~e^{-i\phi_{q}}
\nonumber \\
&\times &\left[ \left( \int y(\phi_{\bar{q}})\tilde{h}_{q}(\phi
_{q},\phi_{\bar{q}})d\phi_{\bar{q}}\right) C(\phi
_{q})+(...)^{T}\right]. \label{hphtotpp}
\end{eqnarray}
In numerical applications, the two-dimensional integrals
over the gauge angles are replaced by a sum over $L_n\times L_p$ points
using the Gauss-Chebyshev
quadrature method \cite{[Har79]}.

\subsection{Canonical representation}

The canonical-basis  single-particle wave functions,
\begin{equation}
\chi_{\mu}({\bm r},\sigma )= \sum_{n} W_{n\mu}~ \psi_{n}({\bm
r},\sigma), \label{canset}
\end{equation}
are defined by the unitary matrix $W$ which diagonalizes the  density matrices,
\begin{equation}
\begin{array}{llr}
   {\sum\limits_{n'} {\rho _{nn'} W_{n'\mu } } } &  =  & v_\mu^2  W_{n\mu } ,  \\
    &  &   \\
   \sum\limits_{n'} \tilde{\rho}_{nn'} W_{n'\mu} &  =  & u_\mu  v_\mu   W_{n\mu },
\end{array}
\end{equation}
where  $v_\mu^2$ are the  occupation probabilities
$0 \leq v_\mu^2 \leq 1$ and $v_\mu^2 + u_\mu^2=1$.
In the canonical representation, the gauge-angle-dependent matrices become
diagonal with the diagonal matrix elements given by:
\begin{eqnarray}
C_{\mu}(\phi_{q}) &=&\frac{e^{2\imath \phi_{q}}}{u_{\mu
q}^{2}+e^{2\imath \phi_{q}}v_{\mu q}^{2}},  \label{eq:4301} \\
\rho_{\mu q}(\phi_{q}) &=&\frac{e^{2\imath \phi_{q}}v_{\mu
q}^{2}}{u_{\mu q
}^{2}+e^{2\imath \phi }v_{\mu q}^{2}},  \label{eq:4303} \\
\tilde{\rho}_{\mu q}(\phi_{q}) &=&\frac{e^{\imath \phi_{q}}u_{\mu
q}v_{\mu
q}}{u_{\mu q}^{2}+e^{2\imath \phi_{q}}v_{\mu q}^{2}},  \label{404}
\end{eqnarray}
and the determinant of matrix $C(\phi_{q})$, needed in
Eq.~(\ref{xphi}), becomes a product of the diagonal values (\ref{eq:4301}).
The use of the  canonical representation significantly
simplifies calculations of the projected fields.

\subsection{Intrinsic average particle number in the HFB+{\VAP} method}\label{Fermi}

The HFB state $|\Phi\rangle$ is a linear combination
of eigenstates $|N\rangle$ of the particle-number operator, i.e.,
\begin{equation}\label{decomp}
|\Phi \rangle = \sum_{N} a_N |N \rangle ,
\end{equation}
where
\begin{equation}
|N\rangle = \frac{P^N |\Phi\rangle}{\sqrt{\langle\Phi|P^N|\Phi\rangle}},
\end{equation}
and $\hat{N}|N\rangle=N|N\rangle$.
The HFB+{\VAP} method is based on the variation of the projected
energy (\ref{eqpnp}), which is the average value of the Hamiltonian on the
state $|N\rangle$, $E^N=\langle N|\hat{H}|N\rangle$.
Obviously, the projected energy does not depend on amplitudes
$a_N$, although the intrinsic average number of particles,
\begin{equation}
\bar{N} = \langle\Phi|\hat{N}|\Phi\rangle = \sum_{N} |a_N|^2N = \mbox{\rm Tr}\rho , \label{intnorm}
\end{equation}
does depend on $a_N$.

The HFB+{\VAP} variational procedure gives, in principle, the same value of the
projected energy independently of the value of $\bar{N}$. This
independence can, however, be subject to numerical instabilities
whenever the amplitude $a_N$, corresponding to the projection on $N$
particles, is small. Therefore, for practical reasons,  one is
interested in keeping the average number of particles $\bar{N}$ as
close as possible to $N$, which guarantees that amplitude $a_N$ is as
large as possible.

In the standard HFB equations (\ref{shfbeq}), the average number of
particles is kept equal to $N$ by adjusting the Lagrange multiplier
$\lambda$. However, in the HFB+{\VAP} approach, $\lambda$ does not
appear in the variational equations (\ref{pshfbeq}), because the
variation of the constant term
$\lambda\langle\Phi|\hat{N}P^N|\Phi\rangle/\langle\Phi|P^N|\Phi\rangle=\lambda N$
equals zero. Therefore, the HFB+{\VAP} equations (\ref{pshfbeq}) do not
allow for adjusting the average particle number $\bar{N}$, which, during the
iteration procedure, may become vary different from $N$. Moreover,
such uncontrolled changes of $\bar{N}$ from one iteration to another
may preclude reaching the stable self-consistent solution.

In order to cope with these problems, one can artificially reintroduce a constant
$\mu$, analogous to the Fermi energy $\lambda$,
into the HFB+{\VAP} equations (\ref{pshfbeq}), i.e.,
\begin{equation}
\left(
\begin{array}{cc}
h^{N}-\mu & \tilde{h}^{N} \\
\tilde{h}^{N} & -h^{N}+\mu
\end{array}
\right) \left(
\begin{array}{c}
\varphi^{N}_{1,k} \\
\varphi^{N}_{2,k}
\end{array}
\right) ={\cal E}_k\left(
\begin{array}{c}
\varphi^{N}_{1,k} \\
\varphi^{N}_{2,k}
\end{array}
\right) ,  \label{pshfbeqmu}
\end{equation}
provided it is equal to zero once the convergence is achieved. With
this modification, the iterations proceed as follows. Suppose that at
a given iteration, condition $\bar{N}$=$N$ is fulfilled. Then, no
readjustment of $\bar{N}$ is necessary, and in the next iteration
one proceeds with $\mu$=0.

Had such an ideal situation continued till the end, a nonzero value
of $\mu$ would have never appeared, and the required solution would
have been found. In practice, this situation never happens, and at
some iteration one finds that $\mbox{\rm Tr}\rho$, i.e., the sum of
norms of the second components $\varphi^N_{2,nk}$ (\ref{denmc3}) of
the HFB+{\VAP} wave functions, is larger (smaller) than $N$. In such
a case, in the next iteration one uses a slightly negative (positive)
value of $\mu$, which decreases (increases) the norms of
$\varphi^N_{2,nk}$,  and decreases (increases) the average particle
number in the next iteration. Since $\mu$ acts in exactly the same
way as the Fermi energy does within the standard HFB method, the 
well-established algorithms of readjusting $\lambda$ can be  used.
Moreover, as soon as the iteration procedure starts to converge, the
non-zero values of $\mu$ cease to be needed, and thus $\mu$ naturally
converges to zero, as required. In practice, we find that the above
algorithm is very useful, and it provides
the same converged solution with any value of
$\bar{N}=N\pm\Delta{N}$, $\Delta{N}$ being a small integer.

\subsection{The cut-off procedure for the contact pairing force}

When using zero-range pairing forces such as the density-dependent
delta force, one has to introduce the energy cut-off \cite{[Dob96]}. Within the
unprojected HFB calculations, a pairing cut-off is introduced by
using the so-called equivalent single-particle spectrum
\cite{[Dob84]}. After each iteration, one calculates an
equivalent spectrum $\bar{e}_{n}$ and corresponding pairing gaps
$\bar{\Delta}_{n}$:
\begin{equation}
\bar{e}_{n}=(1-2P_{n})E_{n}+\lambda,~~~\bar{\Delta}_{n}=2E_{n}\sqrt{P_{n}(1-P_{n})},
\label{ENnbar}
\end{equation}
where $E_{n}$ is the quasiparticle energy,
$\lambda$ is the chemical potential,  and $P_{n}$ denotes the
norm of the lower component of the HFB wave function.
The energy cut-off is practically realized by requesting that
the phase space for the
pair scattering is limited to  those quasiparticle states for which
$\bar{e}_{n}$ is less than the cut-off energy
$\epsilon_{\text{cut}}$ (usually $\epsilon_{\text{cut}}=$ 60\,MeV)
\cite{[Dob01a]}.

Obviously, the above procedure cannot be directly applied to the
HFB+{\VAP} method,  where the intrinsic
quantities, in  particular the `quasiparticle'
energies $E^N_{n}$,  do not have obvious physical meaning.
A reasonable practical prescription for $\epsilon_{\text{cut}}$
can be proposed in terms of  intrinsic ($\phi=0$) HFB fields
$h$ and $\tilde{h}$. After each iteration of
Eq.~(\ref{pshfbeq}), the average quasiparticle energies,
\begin{equation}
 E_n = \left(
\begin{array}{c}
U^N \\
V^N
\end{array}
\right)_n^\dagger \left(
\begin{array}{cc}
h-\lambda & \tilde{h} \\
\tilde{h} & -h+\lambda
\end{array}
\right) \left(
\begin{array}{c}
U^N \\
V^N
\end{array}
\right)_n, \label{newqpe}
\end{equation}
together with
$P_{n}$$\equiv$$P^N_{n}$, give the equivalent energies
(\ref{ENnbar}).   Based on the spectrum of $\bar{e}_{n}$,
the set of quasiparticle states appearing below
the cut-off energy can now be easily defined. At the same time,
the Fermi energy $\lambda$ (as an auxiliary quantity) can
be recalculated in each iteration.

\begin{figure}[htb]
\includegraphics[width=0.48\textwidth]{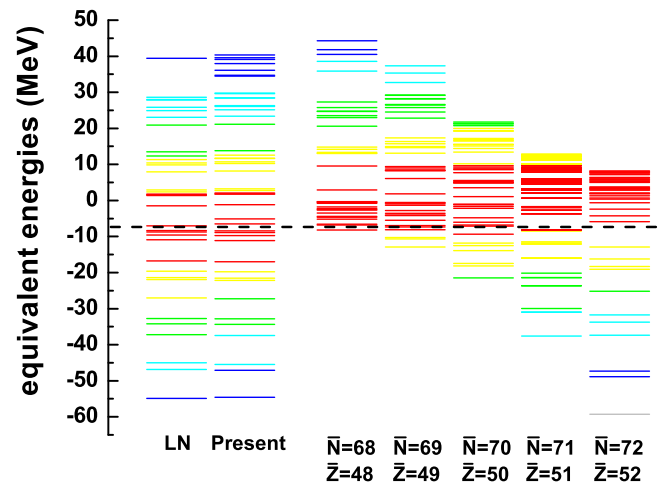}
\caption{\label{fig2}(color online) The neutron equivalent single-particle
energies (\protect\ref{ENnbar}) for $N$=70 and $Z$=50 obtained in the HFB+LN
method (first spectrum), HFB+{\VAP} method using the average quasiparticle
energies $E_n$ (second spectrum) (\protect\ref{newqpe}), and
by using the `quasiparticle' energies
$E^N_k$ calculated in the HFB+{\VAP} method for different values of the intrinsic
neutron $\bar{N}$ and proton $\bar{Z}$ numbers (the
remaining five spectra).
The dashed line indicates the position of the LN Fermi energy $\lambda$.
}
\end{figure}

The results of such a procedure are illustrated in Fig.~\ref{fig2}.
The left-most  spectrum shows the neutron equivalent energies obtained
within the LN method applied to $N$=70 and $Z$=50, and the dashed
line shows the position of the corresponding LN neutron Fermi energy
$\lambda$. For $\bar{e}_{n}<0$, this spectrum is very similar to the
HF bound single-particle energies of this nucleus. Our method, based
on the average quasiparticle energies (\ref{newqpe}), gives almost
identical negative equivalent energies and quite similar positive
ones. In particular, for highly positive equivalent energies, in the
region of the cut-off energy $\epsilon_{\text{cut}}=$ 60\,MeV,
similar continuum quasiparticle states appear in both methods; this
guarantees the correct application of the cut-off procedure. The five
equivalent spectra shown on the right hand side  of Fig.~\ref{fig2} were
calculated directly from the unphysical `quasiparticle' energies
$E^N_{n}$ obtained for several selected values of the intrinsic
particle numbers $\bar{N}$ and $\bar{Z}$. It is obvious that these
spectra (even at $\bar{N}$=70 and $\bar{Z}$=50) bear no resemblance
to the real single-particle spectra and cannot be used to define
the cut-off procedure.

\section{Sample results}
\label{sec5}

To illustrate the Skyrme HFB+{\VAP} procedure, we carried out
calculations for the complete chain of calcium isotopes, from the
proton drip line to the neutron drip line, and for the chain of tin
isotopes with $70\leq{N}\leq90$. We used the Sly4 Skyrme force
parameterization \cite{[Cha98]} and the mixed delta pairing
\cite{[Dob01c],[Dob02c]}. The calculations were performed in the
basis of  20 major HO shells. We took $L$=13 gauge-angle points, and
this practically ensures exact projection for all considered nuclei.
We have found that the HFB+{\VAP} procedure is just $L$-times slower
compared  to the PLN method.

In our standard HFB calculations \cite{[Dob96],[Dob01a]}, the
strength of the pairing force (assumed identical  for protons and
neutrons) is usually adjusted at a given cut-off energy
$\epsilon_{\text{cut}}=60$\,MeV to the experimental value  of the
average neutron gap $\tilde{\Delta}_n$=1.245\,MeV in $^{120}$Sn. In
the present study, we used this procedure to fix the pairing force
for all LN and PLN calculations. However, it is well known that the
PNP method requires another strength  of the pairing force.
Unfortunately, the average pairing gap $\tilde{\Delta}_n$ is not
defined within the {\VAP} approach, and the standard procedure for
adjusting the pairing strength is no longer applicable. In this
study, we adjusted the {\VAP} pairing strength to the total energy of
the $^{44}$Ca nucleus calculated in HFB+PLN. A much more consistent
way of fitting the pairing strength should be based on calculating
the mass differences of the odd-mass and even-even nuclei, all
obtained within the {\VAP} method. We intend to adopt such a
procedure in future applications.

\begin{figure}[htb]
\includegraphics[width=0.48\textwidth]{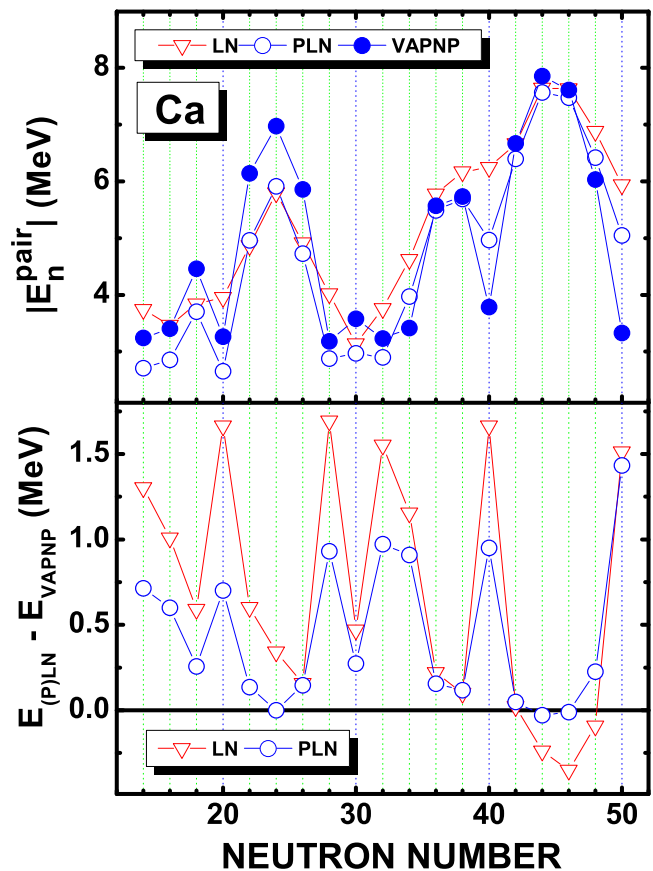}
\protect\caption{\label{fig3} Comparison between the LN, PLN, and {\VAP}
results for the chain of Ca isotopes. The upper panel shows the  neutron
pairing energies while the lower panel shows the total LN and PLN energies
relative to the {\VAP} values.}
\end{figure}

\begin{figure}[htb]
\includegraphics[width=0.48\textwidth]{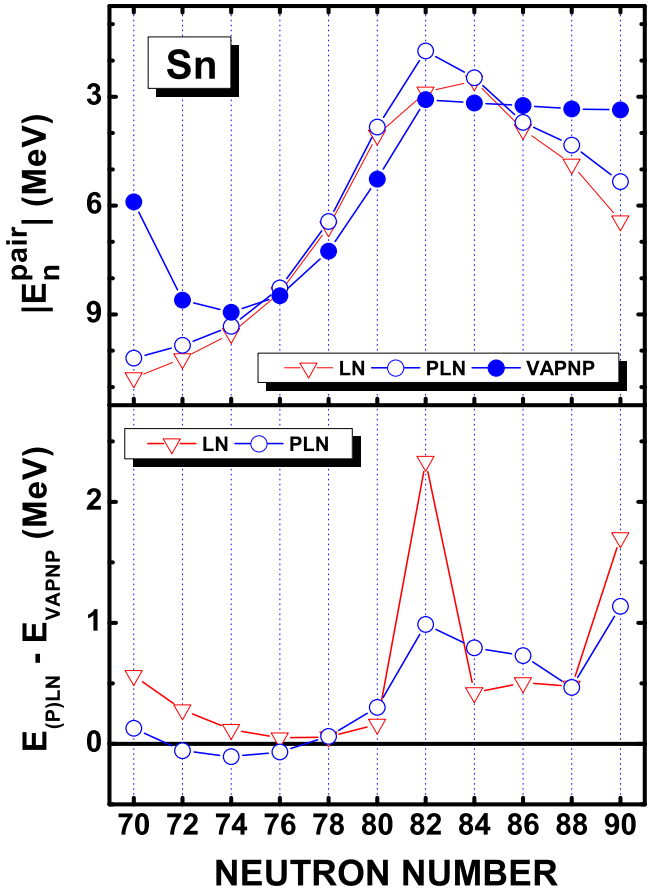}
\protect\caption{\label{fig5}
Similar to Fig.~\protect\ref{fig3}, except for the chain of Sn isotopes.}
\end{figure}

A measure of pairing correlations in a nucleus is the
particle-particle energy (pairing energy) given by the second term
in Eq.~(\ref{EPNPP}). The energy of proton pairing correlations is
about 2--3\,MeV and it changes smoothly with $N$ along the isotopic
chains. On the other hand, the neutron pairing is significantly
affected by the shell structure. As seen in Figs.~\ref{fig3} and
\ref{fig5},  upper panels, the neutron pairing energies obtained
within the LN, PLN, and {\VAP} methods (and with pairing strengths
adjusted as described above) are quite similar to one another.

The lower panels of Figs.~\ref{fig3} and \ref{fig5} show differences
between the total energies obtained in the LN and PLN methods and
those obtained in {\VAP}. The LN or PLN results are fairly close to
{\VAP} for
 mid-shell nuclei, where the neutron pairing correlations are large
and static in character. Near closed shells,
pairing is dynamic in nature, and the LN/PLN results deviate from
those obtained in {\VAP}. For  open-shell nuclei, the PLN
approximation is particularly  good; in 
the calcium isotopes, the deviations from
the HFB+{\VAP} method usually do not exceed 250\,keV. For the
closed-shell nuclei, on the other hand, the LN method is not
appropriate \cite{[Dob93],[Val00],[Ang02]}, and the energy
differences increase to more than 1\,MeV. Figures~\ref{fig3} and
\ref{fig5} also show that the PLN method always leads to a
considerable improvement over LN, often reducing the deviation of the
total energy with respect to {\VAP} by about 1\,MeV.

\begin{figure}[htb] \includegraphics[width=0.48\textwidth]{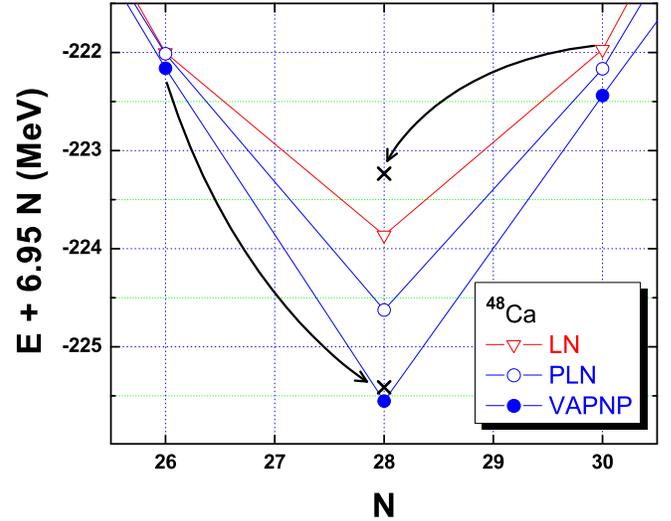}
\protect\caption{\label{fig4} The total binding energy (with respect
to a linear reference) as a function of $N$ for even-even  nuclei
around $^{48}$Ca, calculated in the LN, PLN and {\VAP} methods. Crosses
indicate the PLN results for $^{48}$Ca obtained by projecting from
the LN solutions in neighboring nuclei $^{46}$Ca and $^{50}$Ca
as indicated by arrows.}
\end{figure}

\begin{figure}[htb] \includegraphics[width=0.48\textwidth]{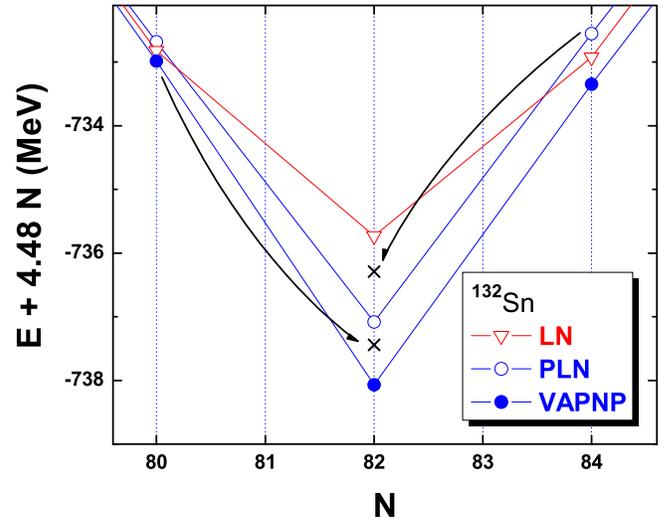}
\protect\caption{\label{fig6}
Similar to Fig.~\protect\ref{fig4}, except for nuclei near $^{132}$Sn.
} \end{figure}

As suggested in Refs.~\cite{[Dob93],[Mag93]},  one can further
improve the PLN approximation around magic nuclei by applying the PNP
to the LN solutions obtained in the neighboring nuclei. This procedure is
illustrated in Figs.~\ref{fig4} and \ref{fig6} for the magic nuclei
$^{48}$Ca and $^{132}$Sn, respectively. It is seen that while the
projection from $^{46}$Ca nicely reproduces the {\VAP} result in $^{48}$Ca, the
approximation fails when projecting from $^{50}$Ca. Similarly,
projection from the LN solution in $^{130}$Sn ($^{134}$Sn) gives a better
(worse) result than the projection of the LN solution obtained in $^{132}$Sn.
We observe a similar pattern of results in other cases near
closed shells; however, the improvement gained by projecting
from isotopes below closed shells is not sufficient to replace
the full {\VAP} calculations at closed shells.

\begin{figure}[htb] \includegraphics[width=0.48\textwidth]{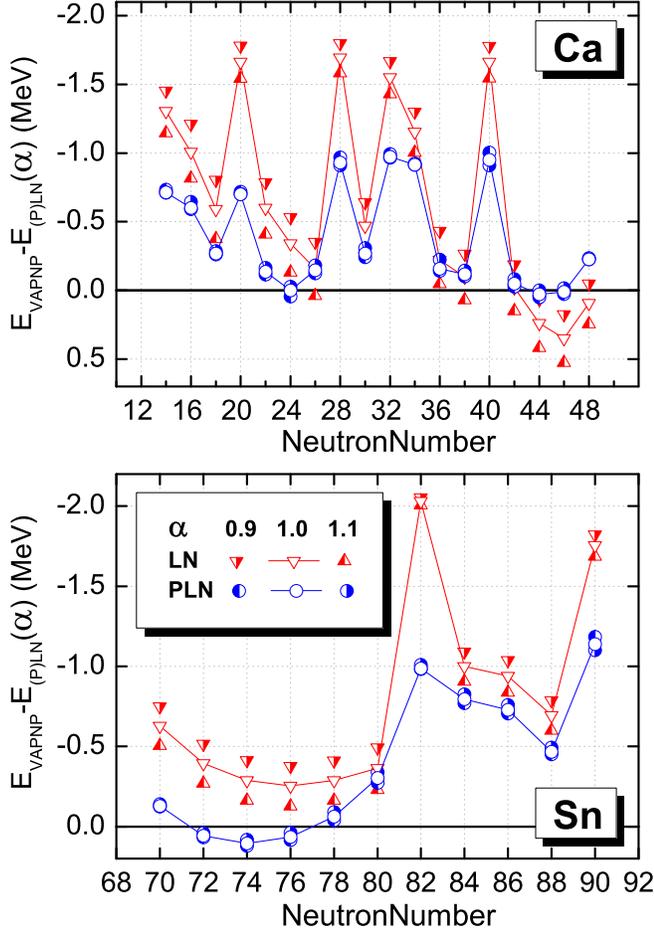}
\protect\caption{\label{fig7} Total LN and PLN energies
relative to the {\VAP} values, calculated in the Ca (upper panel) and Sn
(lower panel) isotopes with the effective pairing strengths
scaled by a factor $\alpha$.}
\end{figure}

In order to discuss the quality of prescription to calculate the LN
parameter $\lambda_2$ presented in Sec.~\ref{LNgen}, we have
repeated all our LN and PLN calculations with the effective pairing
strengths $G'_{\text{eff}}=\alpha G_{\text{eff}}$ scaled by factors
of $\alpha$=0.9 or 1.1 with respect to those given by
Eq.~(\ref{ll4}). In this way, we tested whether our results are
sensitive to this phenomenological prescription. The results obtained
for the chains of Ca and Sn isotopes are shown in Fig.~\ref{fig7}. While 
the LN energies (\ref{enekam})
uniformly depend on the scaling factor $\alpha$,  the PLN
energies are almost independent of the scaling factor. This shows
that the PNP components of the LN states weakly depend on $\lambda_2$
and can be obtained without paying too much attention to the way in
which $\lambda_2$ is calculated. A rough estimate given by our
phenomenological prescription is good enough to obtain reliable PLN
results. On the other hand, deviations between the LN/PLN and {\VAP}
energies depend mostly on the local shell structure and visibly
cannot be corrected by modifications of the prescription used to
calculate $\lambda_2$. In large part, these deviations stem from
the inapplicability of the LN/PLN method to closed-shell nuclei, where
the total energy in function of particle number cannot be well
approximated by the quadratic Kamlah expansion. Altogether, we
conclude that the PLN method gives a fair approximation of the full
{\VAP} results, but fails in reproducing detailed values, especially
near closed shells.

\section{Summary and discussion}
\label{sec6}

In this study, the variation after particle-number projection is discussed
in the context of the nuclear density functional theory. Specifically,
we implement for the first time
the  self-consistent  Skyrme HFB+{\VAP}  formalism. We demonstrate that
the particle-number conserving HFB equations with Skyrme functionals
can be simply obtained from  the standard Skyrme HFB
equations in coordinate space by replacing the intrinsic densities
and currents by their gauge-angle-dependent counterparts.

The calculations are carried for the Ca and Sn isotope chains. The
{\VAP} results are compared with those obtained with the LN and PLN methods.
We demonstrate that the pathological behavior of LN and PLN methods
around closed-shell nuclei can be partly cured by performing
particle-number projection from neighboring open-shell systems. This result
is important in the context of large-scale microscopic mass
calculations, such as those of Ref.~\cite{[Sam04]}.

The restoration of broken symmetries in the density functional
theory causes a number of fundamental questions, mainly
related to the density dependence of the underlying interaction and
to the different
treatment of particle-hole and particle-particle channels
\cite{[Ang01a],[Sto04c]}. These questions and problems will be
discussed in detail in a forthcoming paper.

\begin{acknowledgments}
This work was supported in part by the National Nuclear Security
Administration under the Stewardship Science Academic Alliances
program through the U.S. Department of Energy Research Grant
DE-FG03-03NA00083; by the U.S. Department of Energy
under Contract Nos. DE-FG02-96ER40963 (University of Tennessee),
DE-AC05-00OR22725 with UT-Battelle, LLC (Oak Ridge National
Laboratory), DE-FG05-87ER40361 (Joint Institute for Heavy Ion
Research), DE-FG02-00ER41132 (Institute for Nuclear Theory);
by the Polish Committee for Scientific Research (KBN)
under Contract No.~1~P03B~059~27; and by the Foundation for Polish
Science (FNP).
\end{acknowledgments}


\end{document}